\documentstyle[aps]{revtex}
\def\be{\begin{equation}}
\def\ee{\end{equation}}
\def\bea{\begin{eqnarray}}
\def\eea{\end{eqnarray}}
\begin{document}  

\title{Economical Doublet-Triplet Splitting and Strong Suppression 
of Proton Decay in SO(10)}

\author{ Z. Chacko and Rabindra N. Mohapatra}
\address{\it Department of Physics, University of Maryland,
College Park, MD 20742, USA}

\maketitle

\begin{abstract}
We present a new approach to realizing the Dimopoulos-Wiczek mechanism for
doublet-triplet splitting in supersymmetric SO(10). 
 The method can be used to achieve strong suppression of proton decay
 in a straightforward manner and is relatively economical;
in particular the particle spectrum required for its implementation is
consistent with the constraints from string compactification at Kac-Moody
level two. We construct two examples of realistic models based on this idea. 
These
models have no unwanted flat directions and do not make use of any
nonrenormalizable operators to stabilize the vacuum thereby maintaining
coupling constant unification as a prediction. The first model is 
characterised by a simple discrete symmetry that guarantees naturalness at 
the renormalizable level while also eliminating all potentially dangerous 
R-parity violating effects. The second model predicts the strong 
suppression of Higgsino mediated proton decay. In both models the light MSSM 
doublets
emerge from two different {\bf 10}'s so that there is considerable 
flexibility in constructing realistic fermion masses and CKM angles.
\end{abstract}

\hskip 6cm UMD-PP-99-016
 
\section{Introduction}\hspace{0.5cm} 
One of the attractive features of the minimal supersymmetric extension 
of the standard model (MSSM) is that it leads to unification of the gauge 
couplings at a scale of $M_U$ about $\sim 2\times 10^{16}$ GeV. This is very 
suggestive of the idea that all matter and forces of the standard 
model unify at the scale $M_U$ into a single unifying local 
symmetry\cite{1,2}. The presence of supersymmetry provides the additional 
advantage that the large hierarchy between the weak and the GUT scale can 
be maintained order by order in perturbation theory thanks to the 
non-renormalization theorem of the superpotential.

A key problem of SUSY GUTs is how to split the weak MSSM doublets from the
color triplet fields that accompany them as part of the representation of the
GUT symmetry. This is an essential aspect of SUSY GUTs since both 
coupling constant unification and suppression of proton decay require that
the MSSM doublets $H_u$ and $H_d$ must be at the weak scale whereas the 
triplets which mediate proton decay must have GUT scale mass. This is the 
famous doublet-triplet splitting (DTS) problem.

Another generic problem of SUSY GUT models such as SU(5) and SO(10)
is that they lead to proton decay mediated by colored Higgsino fields.
 In the simplest models the predicted rate is consistent with 
observations only if the Higgsino masses are above the GUT 
scale\cite{nath} and this pushes these models to a very narrow range of 
the allowed parameter space. Alternatively, 
weak suppression can be obtained but only at the expense of large
threshold effects\cite{arno}. It 
is therefore important to seek models that may ameliorate this problem.

There exist several ways to solve the DTS problem. It seems that the one
most appropriate for the SO(10) model is the so called missing vev mechanism
suggested by Dimopoulos and Wilczek(DW)\cite{dimo} where one uses a {\bf 
45}-dim. Higgs which breaks the SO(10) symmetry down to $SU(2)_L\times 
SU(2)_R\times U(1)_{B-L}\times SU(3)_c$ group. The relevant vev is given by:
$<{\bf 45}>=\tau_2\times Diag(a,a,a,0,0)$. If one now couples this multiplet
 to two {\bf 10}-dim. Higgs fields, it is easy to see 
that the MSSM doublet fields in the {\bf 10}'s remain massless. Since 
{\bf 45} is antisymmetric in the SO(10) indices, its coupling must involve two 
different {\bf 10}'s and all four of the standard 
model doublets in the two {\bf 10}'s remain light. Since this is two too 
many, one must add additional couplings which yield a mass term for one 
pair of  doublets. 

 This method has the following serious difficulty. Once one includes 
other multiplets such as the ${\bf 16+\bar{16}}$ to break the $B-L$ 
symmetry the
missing vev pattern essential to implement DTS is destabilized and the 
zeros in the {\bf 45} vev acquire GUT scale vevs; if one forbids the 
coupling of the {\bf 16}'s to the {\bf 45}, then there will be additional 
massless Goldstone states associated with the breaking of a larger global 
symmetry of the superpotential which could upset coupling constant unification. 

This and other difficulties in implementing the Dimopoulos-Wilczek mechanism
 were discussed in a very important paper by 
Babu and Barr\cite{babu} who showed how realistic SO(10) models could be
constructed using this idea. 
They presented a model with a Higgs sector consisting of
three {\bf 45}'s , two {\bf 54}'s, a ${\bf 16+\bar{16}}$ and three {\bf 10}'s
 which could simultaneously 
solve the doublet-triplet splitting as well as the proton decay problem of 
the SO(10) models.  The essence of
their suggestion was that one of the {\bf 45}'s couples to the $ {\bf 
16+\bar{16}}$ pair while a second realizes doublet triplet splitting. The
third {\bf 45} couples to the other two preventing any Goldstone states
from appearing. It also helps suppress proton decay. This model however
leads to unrealistic fermion masses and vanishing CKM angles since a 
single Yukawa coupling determines all of them. The
question of naturalness in the sense of having a symmetry that prevents
all unwanted terms in the superpotential was also not addressed. 
Subsequently alternative models based on the same idea have been 
constructed \cite{babu2}. Unfortunately they also seem to require a similarly 
complicated Higgs sector to achieve DTS. A more economical solution to this 
problem would be attractive from the point of view of simplicity. 

Further motivation for a more straightforward solution is provided by 
string theory.
It is well known that string theories considerably restrict the number and 
nature of the gauge multiplets that survive to low energies. In particular 
the allowed SO(10) reprentations that can emerge as massless multiplets
at the Kac-Moody level of two have been classified \cite{chung} 
\cite{cleaver}. While any number of
vectors ({\bf 10}'s), spinors (${\bf 16+\bar{16}}$'s), and singlets are 
allowed  the number of adjoints ({\bf 45}'s) is restricted to be at most 
two and the number of {\bf 54}'s not more than one. It is of interest to see
whether realistic SO(10) models, and in particular doublet triplet splitting
 can be achieved within these constraints on the particle spectrum.
 
With this in mind there have been some earlier attempts to simplify the 
Higgs content of SO(10) models \cite{babu3}, \cite{barr}, \cite{babu1}.
 However these models make use of 
non-renormalizable operators to realize the desired vacuum which implies
the existence of Higgs fields with mass below the GUT scale, potentially 
upsetting the prediction of gauge coupling unification. In addition some 
of these models have flat directions which are undetermined at the 
classical level; hence their vacuum structure once SUSY is broken is unclear.  
Proton decay in these models is either unsuppressed or weakly suppressed.
 
In this paper, we present a relatively straightforward approach to
realizing the
Dimopoulos-Wilczek mechanism for doublet-triplet splitting in
supersymmetric SO(10). The method is relatively economical; 
in particular the
particle spectrum required for its implementation is
consistent with the above mentioned constraints from string theory.
Specifically the minimal field content consists of 
one {\bf 54}(denoted by $S$), two {\bf 45}'s 
($A, \bar{A}$), two {\bf 10}'s ($H_{1,2}$) and 
a ${\bf 16+\bar{16}}$ pair (denoted by $C+\bar{C}$).
 The essence of our idea is that one {\bf 45} ($\bar A$) acquires the 
Dimopoulos-Wilczek pattern of vevs while the other ($A$)
couples to the 
${\bf 16+\bar{16}}$. Since both these fields couple to the {\bf 54}
there are no unwanted Goldstone modes. We construct two examples of
realistic models based on this scheme. 
 The models have no flat directions (apart from the light 
doublets) and do not use any nonrenormalizable 
terms to determine the desired vacuum, thereby maintaining coupling constant
unification as a prediction. 
The first model is characterised by a simple discrete symmetry which
guarantees naturalness at the renormalizable level while also eliminating
all potentially dangerous R-parity violating effects and is described in 
sec. II. The second 
model predicts the strong suppression of Higgsino mediated proton decay
along with doublet-triplet splitting. This is described in sec. III.
Realistic fermion masses and CKM angles also
arise very easily in both models essentially because
the $H_u$ and $H_d$ of the MSSM emerge from two different {\bf 10}'s as 
suggested in\cite{babu1}. As a result of this the up and the down sector 
mass matrices arise from different Yukawa couplings and therefore the 
CKM angles do not vanish.

The next two sections are devoted to a detailed analysis of the 
supersymmetric vacua of the two models that we present as well as
a demonstration of doublet triplet splitting and proton decay in the case 
of the second model.

\section{Model I:}

 \subsection{Symmetry Breaking Sector}

The part of the superpotential relevant for the breaking of SO(10) down to
the MSSM is given by:
\begin{eqnarray}
W_H= M_S S^2 + \lambda_S  S^3 + M_A [A^2 +\beta
\bar{A}^2]\nonumber\\
 +\lambda_1 S[A^2+\alpha\bar{A}^2] +\gamma Tr S\nonumber\\ + 
2 M_C C\bar{C} - \lambda_C C A \bar{C}  
\end{eqnarray}
where the various fields transform under SO(10) as explained in the 
introduction. We further impose a discrete $Z_4$ symmetry on the theory 
which (as will be clear later) guarantees naturalness at the renormalizable
level.
The transformations of the various fields under  this symmetry are shown in
Table I.
$\gamma$ is the Lagrange multiplier that imposes the trace condition.
We can assume the following vacuum configurations for the various fields:
\begin{eqnarray}
<S>=I_2\times Diag(s_1,s_1,s_1,-\frac{3}{2}s_1,-\frac{3}{2}s_1,)\nonumber\\
<A>=\tau_2\times Diag (a,a,a,b,b)\nonumber\\
<\bar{A}>=\tau_2\times Diag(\bar{a},\bar{a},\bar{a},\bar{b}, 
\bar{b})\nonumber\\
<C_{\nu^c}>=<\bar{C}_{\nu^c}>=v;~~~  
\end{eqnarray}
The equality of the vev's of $C$ and $\bar{C}$ is dictated by the
vanishing of the D-terms at the GUT scale needed to mainatin 
supersymmetry down to the electroweak scale.

Setting the various $F$-terms to zero, we get the following equations. 
For $F_S=0$, we have:
\begin{eqnarray}
5M_Ss_1 -\frac{15}{4}\lambda_S s^2_1 +\lambda_1[a^2-b^2]
+\alpha\lambda_1[(\bar{a}^2-\bar{b}^2]=0
\end{eqnarray}
From the equation of motion for $\bar A$ we get
\begin{eqnarray}
(M_A\beta +\lambda_1 s_1 \alpha) \bar{a}=0\nonumber\\
(M_A\beta -\frac{3}{2}\lambda_1 s_1 \alpha) \bar{b}=0
\end{eqnarray}
The above two equations have three solutions: (i) $\bar{a}=0$ and 
$\bar{b}=0$; (ii)$\bar{a} \neq 0$, which means that $s_1=-\frac{\beta 
M_A}{\lambda_1\alpha}$ and $\bar{b}=0$ and (iii) $\bar{b}\neq 0$ which means
$s_1=\frac{2 M_A\beta}{3\lambda_1\alpha}$ and $\bar{a}=0$. All of these 
are inequivalent vacuua. We will assume that we are in the second one
so that $\bar{b}=0$ and the DW mechanism is realized.

The equations of motion for $A$ yield (using the solution for $s_1$ from
above) 
\begin{eqnarray}
a M_A(1- \frac{\beta}{\alpha})+\frac{1}{2} \lambda_C v^2=0
\end{eqnarray}
\begin{eqnarray}
b M_A(1+ \frac{3\beta}{2\alpha})+\frac{1}{2} \lambda_C v^2=0
\end{eqnarray}
 The $F_C=F_{\bar{C}}=0$ conditions yield
\begin{eqnarray}
M_C +\lambda_C [3a+2 b]=0
\end{eqnarray}

Using the above Eq.(3), (5), (6) and (7), we can determine the values of
$b,a,\bar{a}$ and $v$, in terms of the parameters in the  
superpotential. Note that $\bar A$ has the Dimopoulos-Wilczek
pattern of vevs. Since both this field and the $A$ field
which couples to the {\bf 16}'s couple to the {\bf 54} there are
no unwanted Goldstone modes.

\subsection{Doublet-triplet splitting and Fermion masses}

There are many ways to use the missing vev pattern described above to 
obtain the MSSM doublets from the GUT scale physics. The simplest is to
use the superpotential 
\begin{eqnarray}
W_1 = \lambda_H H_1 \bar{A} H_2 + CC(pH_1+qH_2)+
\bar{C}\bar{C}(p'H_1+q'H_2)
\end{eqnarray}
However we would prefer to
implement the DTS in such a way that it is consistent with the discrete $Z_4$ 
symmetry
that ensures the naturalness of the theory. To this end we introduce an
additional ($\bf{16},\bf{\bar{16}}$) pair, $P$ and $\bar{P}$. We can 
self consistently assume that they have zero vev in the supersymmetric 
limit. Then the part of the superpotential that generates the light MSSM 
doublets $W_{DTS}$ is given below
\begin{eqnarray}
W_{DTS}= \lambda_H H_1 \bar{A} H_2 + 
CP(pH_1+qH_2)+\bar{C}\bar{P}(p'H_1+q'H_2) + \lambda_{P}P\bar{A}\bar{P}  
\end{eqnarray}
The role of the various terms in $W_{DTS}$ is not hard to understand.
At the GUT scale, our model has four up-type and four down type 
MSSM doublets; of these the doublets belonging to $C,\bar{C}$ become 
superheavy due to the Higgs sector described in the previous section 
leaving us with three
up and down Higgs doublets i.e. 
($H_{1u}, H_{2u}$, $\bar{P}_u$) and ($H_{1d}, 
H_{2d}, P_d$). It is clear from Eq.(9) that $P_d$ and one linear 
combination of $H_{1u}$ and $H_{2u}$ pair up to 
become superheavy and $\bar{P}_u$ and a linear combination of 
$H_{1d}$ and $H_{2d}$ also pair up and decouple. 
The light MSSM Higgs doublets therefore are the orthogonal linear combinations.
On the other hand, the triplets present in $H_{1,2}$ become superheavy 
due to the $H_1H_2A$ term in the $W_{DTS}$, where as those in $P$ and 
$\bar{P}$ become superheavy by the similar term. The triplets in $C$ and 
$\bar{C}$ also become superheavy due to the Higgs sector described in
the previous section. Thus 
there are no light triplets below GUT scale. 
This method of using the {\bf 16}'s to split doublets was first noted in
Ref.\cite{babu1} and was subsequently used in \cite{barr}.It is 
straightforward to verify 
that the superpotential for the Higgs fields $W_H +W_{DTS}$ has the most 
general form compatible with the gauge and discrete symmetries at the
renormalizable level.

In order to discuss the fermion masses in this model, we must write down the
Yukawa part of the superpotential which is invariant under the 
$Z_4$ symmetry. Using the transformation properties in of the fields under 
these symmetries given in table I, we get:
\begin{eqnarray}
W_Y= h_{ab}\psi_a\psi_b H_2 + h'_{ab}\psi_a\psi_b H_1 \nonumber\\
+h''_{ab}\psi_a\psi_b A H/M +...
\end{eqnarray}
Using this one can construct realistic fermion masses. 
Note that the Yukawa matrices are arbitrary in 
generation space which allows for many more possibilities for mass textures.

It is well known that in SO(10) models where the $B-L$ symmetry is broken by
{\bf 16} vevs, operators like $\psi\psi\psi C/M_{Pl}$ lead after symmetry 
breaking to baryon and lepton number violating (R-parity violating) terms
with strengths of order $10^{-2}$, in strong disagreement with the 
existing limits on such couplings from proton decay etc.\cite{bhat}. It is 
worth pointing out that the same discrete symmetry that led to a natural 
realization of the doublet-triplet splitting also forbids all
dangerous R-parity violating terms. The lowest order term allowed by the 
symmetry is of the form $\psi^6 C^2/M^5_{P}$. This leads to a generic
sixth order term in quark fields such as 
$u^cu^c\tilde{d^c}\tilde{d^c}\tilde{d^c}\tilde{d^c}$ 
with a strength of order $10^{-58}$ GeV$^{-3}$ and has no observable effect.

In this model, in the limit of $p,p',q, q',h'_{ab}=0$, Higgsino mediated 
contributions to proton decay vanish. However in this limit, CKM angles 
also vanish and there is an extra pair of MSSM doublets. Therefore, it is 
possible to get additional suppression over the minimal SU(5) prediction for 
proton decay as long as the extra doublet pair is slightly below the GUT 
scale. This would generate new threshold corrections which should be taken 
into account in the discussion of unification.

 \section{Model II:}

\subsection{Symmetry Breaking Sector}

As demonstrated in \cite{babu}, strong suppression of proton decay can be
realized in supersymmetric SO(10) if a $\bf{45}$ has the complimentary 
pattern of vev's to the Dimopoulos-Wilczek pattern, ie 
$\tau_2\times Diag (0,0,0,a,a)$. In this section we present the sector of 
the model which breaks SO(10) down to the MSSM. We will show that in this
process, this pattern of vevs is generated for the field $A$. The field
$\bar A$ will be shown to retain the Dimopoulos-Wilczek form, thereby
realizing doublet-triplet splitting.
The relevant part of the superpotential is given by
\begin{eqnarray}
W^{1}_{SB} =~M_1S^2 + \lambda S^3 + SA\bar{A} + 
\frac{3}{2} YA\bar{A}\nonumber \\ + \gamma Tr S
+M_0 A^2 -\beta CA\bar{C} + 2M_2 C\bar{C} 
\end{eqnarray}
where $Y$ is an SO(10) singlet field, and the other fields transform 
as before. We have scaled all fields so as to set some couplings 
to one.
We can assume the following vacuum configurations for the various fields:
\begin{eqnarray}
<S>=I_2\times Diag(s_1,s_2,s_3,s_4,s_5)\nonumber\\
<A>=\tau_2\times Diag (a_1,a_2,a_3,a_4,a_5)\nonumber\\
<\bar{A}>=\tau_2\times Diag(\bar{a_1},\bar{a_2},\bar{a_3},\bar{a_4}, 
\bar{a_5})\nonumber\\
<C_{\nu^c}>=<\bar{C}_{\nu^c}>=v
\end{eqnarray} 

The equations for the vanishing $F$-terms at the GUT scale are:
\begin{eqnarray}
\Sigma_i a_{i}\bar{a_{i}}=0
\end{eqnarray}
\begin{eqnarray}
s_i\bar{a_{i}}+\frac{3}{2}y\bar{a_{i}} + 2M_0 a_{i} +\beta v^2=0
\end{eqnarray}
\begin{eqnarray}
3\lambda s^2_i+2M_1 s_i + \bar{a_{i}}a_{i} +\gamma=0
\end{eqnarray}
\begin{eqnarray}
s_ia_{i} +\frac{3}{2} y a_{i}=0
\end{eqnarray}
\begin{eqnarray}
M_2 +\beta \Sigma_i a_{i}=0
\end{eqnarray}
All vev's are given in obvious notation.
We are interested in the ground state for which $a_{4}=a_{5}\neq 0$,
$a_{\alpha}=0$ (for $\alpha=1,2,3$), $\bar{a_{\alpha}}\neq 0$ and 
$\bar{a_{4}}=\bar{a_{5}}=0$. Let us start with the first two equations and use 
the  
$F$-term conditions above to show that the second two equations follow. From 
the above equations it is easy to see 
that the first two conditions (i.e. $a_{4,5}\neq 0$ and $a_{\alpha}=0$) 
imply that $s_{4}=s_{5}= -\frac{3}{2} y$ leaving $s_{\alpha}$ undetermined.
Eq. (15) and the tracelessness of $S$ then imply that $s_{\alpha}$ are all 
equal to $y$. Using Eq.(15), we get $a_{4}=a_{5}=-\frac{\beta v^2}{2M_0}$. 
Using Eq. (14) and (15), it is easy to see that 
$\bar{a_{4}}=\bar{a_{5}}=0$ and all $\bar{a_{\alpha}}$ are equal as desired. 
The
Equation (15) for $F_S=0$ then leads to two solutions for $s_i$ whose
sum is equal to $-\frac{2M_1}{3\lambda}$ thus solving for $y$ and $s_i$. 
The Eq.(17) for $F_C=0$ then determines $a_{4}$ and thus $v$; Eq. (14) 
then determines $\bar{a}$. Hence $A$ and $\bar{A}$ have the desired 
pattern of vevs 
$\tau_2\times Diag (0,0,0,a,a)$ and $\tau_2\times Diag (\bar{a},\bar{a},
\bar{a},0,0)$ respectively.

The model as we have presented above has a pair of massless singly charged
fields which arise from the {\bf (1, 3, 1)} (under 
$SU(2)_L\times SU(2)_R\times SU(4)_c$) component of the $\bar{A}$ field.
This problem however can be easily cured without effecting the rest of
the discussion by introducing an extra pair of ${\bf 16 + \bar{16}}$
(denoted by ${\bf P+\bar{P}}$ as before) and a pair of singlet fields
${\bf T +{\bar{T}}}$. We then add to the above superpotential the following
piece:
\begin{eqnarray}
W^{(3)}= C\bar{A}\bar{P} +\bar{C}\bar{A} P +CT\bar{P} + \bar{C}\bar{T}P
+M'P\bar{P}
\end{eqnarray}
Note that minima equations for $T$ and $\bar{T}$ imply that $P$ and
$\bar{P}$ have vanishing vev's but $T$ and $\bar{T}$ have nonzero vevs.
It is then easy to see that the massless fields of $\bar{A}$ are now lifted
to GUT scale as they pair up with right handed doublets from $P$ and
$\bar{P}$.

\subsection{Doublet-Triplet spltting, Strong Suppression of Proton Decay 
and Fermion Masses in Model II}

It was pointed out in \cite{babu} that by increasing the number of {\bf 
10}-dim. representations to three, one can obtain a strong suppression 
of proton decay, if the following superpotential ($W_{BB}$) is chosen:
\begin{eqnarray}
W_{BB}= H_1\bar{A}H_2 + H_2 A H_3 +M_3 H^2_3
\end{eqnarray}
The problem with this model lies however with the pattern of fermion masses
it leads to. It is easy to see that the light MSSM doublets in this model 
belong to the same {\bf 10}; as a result, one gets vanishing quark mixing 
angles. This problem is solved if we have four {\bf 10}-dim. fields
$H_{1,2,3,4}$ and we choose the following superpotential to implement 
both doublet-triplet splitting and achieve strong suppression of proton 
decay:
 \begin{eqnarray}
W_{DTS}= H_1 \bar{A} H_2 + (H_1 + H_2) A (H_3 + H_4) + H_3 \bar{A} H_4
\nonumber\\
+CCH_1 + \bar{C}\bar{C}H_3
\end{eqnarray}
From $W_{DTS}$ it is clear that there is exactly one pair of massless MSSM 
doublets. This can be seen more explicitly by writing down the mass 
matrix for the doublet fermions in the {\bf 10}'s, the ${\bf C+\bar{C}}$
and the ${\bf P+\bar{P}}$:
\begin{eqnarray}
M_D=\left(\begin{array}{cccccc} H_{1u} & H_{2u} & H_{3u} & H_{4u} & \bar{C}_u
& \bar{P}_u
\end{array}\right)\left(\begin{array}{cccccc}
0 & 0 & a & a & v & 0\\
0 & 0 & a & a & 0 & 0\\
a & a & 0 & 0 & 0 & 0\\
a & a & 0 & 0 & 0 & 0 \\
0 & 0 & v & 0 & 2M_2 & 12 \bar{a}\\
0 & 0 & 0 & 0 & 12\bar{a} & M'\end{array}\right)\left(\begin{array}{c} 
H_{1d} \\H_{2d} \\ H_{3d} \\ H_{4d} \\ C_d \\ P_d \end{array}\right)
\end{eqnarray}
where we have denoted the vev of $A$ by $a$ and that of $\bar{A}$ by
$\bar{a}$. We remind the reader that $<A>= \tau_2 \times Diag(0,0,0,a,a)$
and $<\bar{A}>= Diag (\bar{a},\bar{a}, \bar{a}, 0, 0)$.
The light doublets are then given by
 \begin{eqnarray}
H_u~ =~[ (H_3)_u- (H_4)_u]/\sqrt{2} \nonumber\\
H_d ~=~[(H_1)_d - (H_2)_d]/\sqrt{2}
\end{eqnarray}
To see the color triplet masses, we write their mass matrix as follows:
\begin{eqnarray}
M_T~=~\left(\begin{array}{cccccc} \xi_1 & \xi_4 & \xi_2 & \xi_3 & \xi_{C}
&\xi_P \end{array} \right)\left(\begin{array}{ccccccc}
\bar{a} & 0 & 0 & 0 & 0 & 0 \\
0 & \bar{a} & 0 & 0 & 0 & 0 \\
0 & 0 & \bar{a} & 0 & 0 & 0 \\
0 & 0 & 0 & \bar{a} & v & 0 \\
0 & 0 & v & 0 & 4M_2 & 8\bar{a} \\
0 & 0 & 0 & 0 & 8\bar{a} & M' \\
\end{array}\right)\left(\begin{array}{c}
\bar{\xi}_2 \\ \bar{\xi}_3 \\ \bar{\xi}_1 \\ \bar{\xi}_4 \\
\bar{\xi}_{\bar{C}}\\ \bar{\xi}_{\bar{P}}\end{array} \right)
\end{eqnarray}
First it is clear from this equation that all color triplets have GUT 
scale mass. Moreover note that $\xi_1$ and ${\xi}_4$ do not connect
to $\bar{\xi}_1$ and $\bar{\xi}_4$. As a result, if
we choose the fermion Yukawa couplings to be of the form
\begin{eqnarray}
W_Y= h_u \psi\psi H_1 + h_d \psi\psi H_4
\end{eqnarray}
there are no dimension five Higgsino mediated proton decay operators 
induced by the color triplet fields. 
Furthermore, we the see that the up and the down type fermion masses arise 
from two different couplings
and we therefore have freedom to construct realistic fermion mass matrices.
We can correct for the bad mass relation $m_e/m_{\mu}=m_d/m_s$ by adding
higher dimensional terms of the form $\psi\psi A\bar{A} H_1$ etc as 
explained in \cite{babu1}.

It is conceivable that the form of the superpotential or some 
generalization of it is dictated by a set of discrete symmetries; 
however, we have not made any attempt to find them here.

\section{Conclusion}
In conclusion, we have presented a method of implementing doublet triplet
splitting in supersymmetric SO(10) in a simple and economical way using the
field content dictated by fermionic compactification of superstrings.
Two realistic models have been constructed using this idea. The models have  
no unwanted flat directions, maintain coupling constant unification as a 
prediction and give rise to realistic fermion masses. Strong suppression of
Higgsino mediated proton decay emerges as a prediction of one of the models. 

 This work is supported by the National Science Foundation 
under grant no. PHY-9802551.

\begin{table}[htb]
\begin{center}
\[
\begin{array}{|c||c||c|} \hline
Fields &  Z_4 \\ \hline
H_1 &  iH_1 \\
H_2 &  iH_2 \\
A & A \\
\bar{A} & -\bar{A} \\
C & C  \\
\bar{C} & \bar{C} \\
\psi_a &  e^{\frac{i\pi}{4}}\psi_a \\
P & -iP \\
\bar{P} & -i\bar{P}\\
\hline
\end{array}
\]
\end{center}
\caption{Transformation of the different fields under the discrete symmetry
$ Z_4$.}

\end{table}

\end{document}